\documentstyle[]{article}

\begin{document}

\title{QUANTUM SPIN CHAINS
       \protect\\
       WITH REGULARLY ALTERNATING BONDS AND FIELDS}

\author{Oleg Derzhko\\
        Institute for Condensed Matter Physics,\\
        1 Svientsitskii Street, L'viv-11, 79011, Ukraine}

\maketitle

\begin{abstract}
We consider the spin-$\frac{1}{2}$ $XY$ chain in a transverse field
with regularly varying exchange interactions and on-site fields.
In two limiting cases of the isotropic ($XX$)
and extremely anisotropic (Ising) exchange interaction
the thermodynamic quantities are calculated rigorously
with the help of continued fractions.
We discuss peculiarities of the low-temperature magnetic properties
and a possibility of the spin-Peierls instability.
\end{abstract}


\vspace{2mm}


\vspace{5mm}

\renewcommand\baselinestretch{1.5}
\large\normalsize

It is generally known
that there are a number of one-dimensional spin models
for which in the translationally invariant (uniform) case
an exact calculation of the thermodynamic quantities is possible.
To this class belong
the spin-$\frac{1}{2}$ $XY$ models
proposed by Lieb, Schultz and Mattis \cite{[1]}.
At present
there are relatively large number of magnetic and ferroelectric compounds
which may be described by these models.
On the other hand,
due to the progress in material sciences
a study of the properties
of quantum spin chains with regularly varying parameters
(e.g., exchange interactions or on-site fields)
attracts much interest during last years.
As an example
one may mention the layered materials
with the well-pronounced one-dimensional character in their properties
known as superlattices,
which can be viewed as chains with regularly varying exchange interactions
(see, e.g., \cite{[2]}).
In what follows
we show how the thermodynamic properties
of the regularly alternating one-dimensional
spin-$\frac{1}{2}$ $XY$ model in a transverse field
can be examined rigorously
in two limiting cases
of the isotropic exchange interaction
(transverse $XX$ chain)
and
of the extremely anisotropic exchange interaction
(transverse Ising chain).
We discuss briefly
the effects of regular alternation on the magnetic properties
of the chains
at zero temperature.
We also comment on a possibility
of the spin-Peierls instability in the considered chains.

We consider a ring of $N\to\infty$ spins $1\over 2$
described by the Hamiltonian
\begin{eqnarray}
H=\sum_{n}\Omega_ns_n^z
+2\sum_{n}\left(I_n^xs_n^xs_{n+1}^x+I_n^ys_n^ys_{n+1}^y\right)
\end{eqnarray}
concentrating on two limiting cases
1) of the isotropic exchange interaction
$I_n^x=I_n^y=I_n$
(transverse $XX$ chain)
and
2) of the extremely anisotropic exchange interaction
$I_n^x=I_n$, $I_n^y=0$
(transverse Ising chain).
We assume
that the exchange interactions $I_n$
and the on-site (transverse) fields $\Omega_n$
vary regularly along the chain with period $p$
($N$ is a multiple of $p$),
i.e.,
$I_1\Omega_1 I_2\Omega_2 \ldots I_p\Omega_p
I_1\Omega_1 I_2\Omega_2 \ldots I_p\Omega_p\ldots\;$.
To calculate the thermodynamic quantities of the spin model
we map it onto a system of spinless (Jordan-Wigner) fermions
with the Hamiltonian \cite{[1]}
\begin{eqnarray}
H=\sum_n\Omega_n\left(c_n^+c_n-\frac{1}{2}\right)
\nonumber\\
+\sum_n
\left(
\frac{I_n^x+I_n^y}{2}
\left(c^+_nc_{n+1}-c_nc^+_{n+1}\right)
+
\frac{I_n^x-I_n^y}{2}
\left(c^+_nc_{n+1}^+-c_nc_{n+1}\right)
\right).
\end{eqnarray}
The Hamiltonian (2) can be brought into the diagonal form
$H=\sum_k\Lambda_k\left(\eta^+_k\eta_k-\frac{1}{2}\right)$
where $\Lambda_k$ is the elementary excitation energy
and $\eta^+_k$, $\eta_k$ are Fermi operators.
Obviously,
the diagonalization of the Hamiltonian
for $p>1$ becomes not trivial.

To find the thermodynamic quantities of the spin model
it is sufficient to know either
$\rho(E)=\frac{1}{N}\sum_k\delta(E-\Lambda_k)$
or
$R(E^2)=\frac{1}{N}\sum_k\delta(E^2-\Lambda_k^2)$
since the Helmholtz free energy per site
can be written as follows
\begin{eqnarray}
f=-\frac{1}{\beta}
\int_{-\infty}^{\infty}{\mbox{d}}E\rho(E)\ln\left(2\cosh\frac{\beta E}{2}\right)
=-\frac{2}{\beta}
\int_{0}^{\infty}{\mbox{d}}EER(E^2)\ln\left(2\cosh\frac{\beta E}{2}\right).
\end{eqnarray}

Consider first the transverse $XX$ chain.
We introduce the one-fermion Green functions
$G_{nm}=G_{nm}(E)$
which satisfy the set of equations
\begin{eqnarray}
(E-\Omega_n)G_{nm}-I_{n-1}G_{n-1,m}-I_nG_{n+1,m}=\delta_{nm}
\end{eqnarray}
and yield the required density of states
$\rho(E)
=\mp\frac{1}{\pi N}\sum_n{\mbox{Im}}G_{nn}^{\mp}$,
$G_{nm}^{\mp}=G_{nm}(E\pm{\mbox{i}}\epsilon)$,
$\epsilon\to+0$.
One immediately notes that the continued fraction representation
for the diagonal Green functions which follows from (4)
\begin{eqnarray}
G_{nn}=\frac{1}{E-\Omega_n-\Delta_n^--\Delta_n^+},
\\
\Delta_n^-=\frac{I_{n-1}^2}{E-\Omega_{n-1}
-\frac{I_{n-2}^2}{E-\Omega_{n-2}-_{\ddots}}},
\;\;\;
\Delta_n^+=\frac{I_n^2}{E-\Omega_{n+1}
-\frac{I^2_{n+1}}{E-\Omega_{n+2}-_{\ddots}}}
\nonumber
\end{eqnarray}
is extremely useful
in the case of regularly varying parameters $I_n$, $\Omega_n$
since $\Delta^-_n$, $\Delta_n^+$ in (5) become periodic
and can be evaluated exactly by solving quadratic equations.

We cannot proceed in the described manner
for the transverse Ising chain
because of terms $c^+_nc^+_{n+1}$, $c_nc_{n+1}$ in (2).
However, from \cite{[1]} we know
that to find the (real) coefficients
$g_{kn}=\frac{1}{2}\left(\Phi_{kn}+\Psi_{kn}\right)$,
$h_{kn}=\frac{1}{2}\left(\Phi_{kn}-\Psi_{kn}\right)$
of the canonical transformation
$\eta_k=\sum_n\left(g_{kn}c_n+h_{kn}c_n^+\right)$
which diagonalizes the Hamiltonian (2)
one should solve the set of equations
\begin{eqnarray}
\Omega_{n-1}I_{n-1}\Phi_{k,n-1}
+\left(\Omega_n^2+I_{n-1}^2-\Lambda_k^2\right)\Phi_{kn}
+\Omega_nI_n\Phi_{k,n+1}=0
\end{eqnarray}
(and a similar set of equations for $\Psi_{kn}$).
Eq. (6) coincides with a set of equations
describing displacements of particles
in a nonuniform harmonic chain with nearest neighbour interactions
and $\Lambda_k$ plays a role of oscillator frequency.
We can find the distribution
of oscillator frequencies squares $R(E^2)$
from the relation
$R(E^2)=\mp\frac{1}{\pi N}\sum_n {\mbox{Im}} \Gamma_{nn}^{\mp}$,
$\Gamma_{nm}^{\mp}=\Gamma_{nm}(E^2\pm{\mbox{i}}\epsilon)$,
$\epsilon\to+0$
where the Green functions
$\Gamma_{nm}=\Gamma_{nm}(E^2)$
satisfy the set of equations
\begin{eqnarray}
\left(E^2-\Omega_n^2-I^2_{n-1}\right)\Gamma_{nm}
-\Omega_{n-1}I_{n-1}\Gamma_{n-1,m}
-\Omega_nI_n\Gamma_{n+1,m}=\delta_{nm}.
\end{eqnarray}
From (7) the following representation for $\Gamma_{nn}$ may be derived
\begin{eqnarray}
\Gamma_{nn}
=\frac{1}{E^2-\Omega_n^2-I_{n-1}^2-\Delta_n^--\Delta_n^+},
\\
\Delta_n^-=\frac{\Omega_{n-1}^2I_{n-1}^2}
{E^2-\Omega_{n-1}^2-I_{n-2}^2-\frac{\Omega_{n-2}^2I_{n-2}^2}
{E^2-\Omega_{n-2}^2-I^2_{n-3}-_{\ddots}}},
\nonumber\\
\Delta^+_n=\frac{\Omega_n^2I_n^2}
{E^2-\Omega^2_{n+1}-I_n^2-\frac{\Omega_{n+1}^2I_{n+1}^2}
{E^2-\Omega^2_{n+2}-I_{n+1}^2-_{\ddots}}}.
\nonumber
\end{eqnarray}
Again for a periodic sequence of parameters $I_n$, $\Omega_n$
the continued fractions
$\Delta_n^-$, $\Delta_n^+$ in (8) become periodic
and can be evaluated exactly
by solving quadratic equations.

To illustrate how the described approach works
we consider the chains of period 1 (uniform chains).
Then for the transverse $XX$ chain one finds
\begin{eqnarray}
\Delta_n^-=\Delta_n^+
=\frac{1}{2}\left(E-\Omega\pm\sqrt{\left(E-\Omega\right)^2-4I^2}\right),
\nonumber\\
\rho(E)=\left\{
\begin{array}{ll}
\frac{1}{\pi}\frac{1}{\sqrt{-\left(E-\Omega-2I\right)\left(E-\Omega+2I\right)}},
& {\mbox{if}} \left(E-\Omega\right)^2<4I^2,\\
0, & {\mbox{otherwise}}.
\end{array}
\right.
\end{eqnarray}
For the transverse Ising chain the corresponding results read
\begin{eqnarray}
\Delta_n^-=\Delta_n^+
=\frac{1}{2}\left(E^2-\Omega^2-I^2
\pm\sqrt{\left(E^2-\Omega^2-I^2\right)^2-4\Omega^2I^2}\right),
\nonumber\\
R(E^2)=\left\{
\begin{array}{ll}
\frac{1}{\pi}
\frac{1}{\sqrt{-\left(E^2-\left(\Omega-I\right)^2\right)
\left(E^2-\left(\Omega+I\right)^2\right)}},
& {\mbox{if}} \left(E^2-\Omega^2-I^2\right)^2<4\Omega^2I^2,\\
0, & {\mbox{otherwise}}.
\end{array}
\right.
\end{eqnarray}
Formulas (3), (9), (10) reproduce the long known results,
however,
the described approach
(formulas (3), (5), (8))
yields new results
for the chains of longer periods.
These calculations are more lengthy
(although transparent)
and therefore are omitted here (see \cite{[3],[4]}).

Let us discuss the effects caused by regular nonuniformity
which can be studied rigorously
within the framework of the elaborated  approach.
We start from the transverse $XX$ chain.
Regular alternation of the Hamiltonian parameters
leads to a splitting of the fermion band (see (2))
into several subbands.
Since the (transverse) magnetization per site is
$m=-\frac{1}{2}
\int_{-\infty}^{\infty}{\mbox{d}}E\rho(E)\tanh\frac{\beta E}{2}$
one immediately observes the plateaus
in the dependence $m$ vs. $\Omega$
at zero temperature $\beta\to\infty$
(we assume $\Omega_n=\Omega+\Delta\Omega_n$)
which are accompanied by the square-root singularities
in the zero-temperature dependence
(transverse) susceptibility $\chi$ vs. $\Omega$.
The number of plateaus does not exceed the period of alternation,
the characteristic fields at which the plateaus start and end up
(and the singularities of $\chi$ occur)
can be easily calculated from the density of states $\rho(E)$.
Consider further the case of modulated exchange interactions
$\vert I_{1,2}\vert=\vert I\vert(1\pm\delta)$
($0\le\delta\le 1$ is the dimerization parameter)
and constant field $\Omega$.
One may easily convinced himself
that the ground-state energy per site
$e_0=-\frac{1}{2}\int_{-\infty}^{\infty}{\mbox{d}}E\rho(E)\vert E\vert$
for $\vert\Omega\vert<2\vert I\vert$
decreases sufficiently rapid
with increasing of $\delta$
to provide a spin-Peierls dimerization.
Moreover, the ground-state energy
of the chains having longer periods of exchange interaction modulation
obtained with the help of continued fractions
allows one
to examine the stability of the transverse $XX$ chain
with respect to more complicated lattice distortions.

Let us pass to the transverse Ising chain.
Considering the case of modulated exchange interactions
(e.g., of period 2)
and constant field
and using the strong-coupling approach arguments
(i.e., assuming the smallest interaction to be equal to $0$)
\cite{[5]}
one immediately notes
that no magnetization plateaus should be expected.
However,
using the continued fraction approach results
one may show
that a logarithmic singularity of $\chi$ may occur
if $\Omega=\pm I$ ($p=1$),
$\Omega_1\Omega_2=\pm I_1I_2$ ($p=2$) etc.
As a result
the transverse Ising chain may exhibit `plateaus'
in the dependence $m$ vs. $\Omega$
for sufficiently large deviations of fields $\Delta\Omega_n$.
Moreover,
in contrast to $XX$ chain,
the number of singularities of $\chi$
(and the number of `plateaus')
for given $p$
depends on the values of Hamiltonian parameters.
The direct calculation of the ground-state energy
of the dimerized Ising chain in a constant transverse field
shows that no spin-Peierls instability should be anticipated
that is in agreement with general arguments
about the Peierls mechanism.
Note, however,
that the transverse Ising chain
is unitarily equivalent
to the anisotropic $XY$ chain without field \cite{[6]}.
The latter model
for a special choice of parameters
may become the isotropic $XY$ model
for which, as we know from the rigorous calculation,
the spin-Peierls instability does occur.

To summarize,
we have discussed an application of continued fractions
for calculation of the thermodynamic quantities
of the regularly alternating spin-$\frac{1}{2}$ transverse $XX$ and Ising chains.
We have briefly discussed the magnetic properties of such chains
as well as their stability with respect to a lattice distortion
due to the spin-Peierls mechanism.

\end {document}